\documentclass[runningheads]{llncs}
\usepackage[T1]{fontenc}
\usepackage{graphicx}
\usepackage{parskip}

\usepackage[colorlinks=true,allcolors=blue,breaklinks,draft=false]{hyperref}

\usepackage{orcidlink}
\usepackage{float}
\usepackage{fancyvrb}
\floatstyle{plain}
\newfloat{listing}{tbp}{lol}
\newcommand{\startmyfigspace}{\vspace{-2ex}}
\newcommand{\myfigspaceend}{\vspace*{-2ex}}
\newcommand{\startmycapspace}{\vspace*{-2ex}}

\definecolor{darkgreen}{RGB}{1,50,32}

\newif\ifkeepcomments
\keepcommentstrue

\ifkeepcomments
\newcommand{\seth}[1]{{{\bf\Large\color{purple}\footnote{\color{purple}Seth: #1}}}}
\newenvironment{iseth}{%
  \color{red}%
}{}
\else
\newcommand{\seth}[1]{\relax}
        {\expandafter\comment%
}
        {\expandafter\endcomment%
}
\fi

\begin{document}
\title{Personalized Assessments from Personal Artifacts}
%
%
\author{Yufan Zhang\orcidlink{0009-0004-3791-7311}
\and Jaromir Savelka\orcidlink{0000-0002-3674-5456}
\and Seth Copen Goldstein\orcidlink{0000-0003-1512-0446}
\and Majd Sakr\orcidlink{0000-0001-5150-8259}
}
\authorrunning{Zhang et al.}
%
\institute{Carnegie Mellon University, Pittsburgh, PA 15213, USA\\
\email{\{yufanz,jsavelka,sethg,msakr\}@andrew.cmu.edu}}
\maketitle              
\begin{abstract}

The rapid development and popularization of AI-enabled coding agents have meant software engineering students and professionals cannot be assumed to understand their own code, which risks academic integrity and professional accountability. We developed a method called Personalized Probing Puzzles ($p^3$) to evaluate students' understanding of their own code, and tested $p^3$ in a graduate-level cloud computing course. Our pilot study shows that $p^3$ can help identify potential gaps in students' understanding of their own code. The puzzles are automatically generated, asynchronously administered, and finished in minutes. Future work is needed to correlate puzzle results with code understanding and to embed $p^3$ in a professional code review process.

\end{abstract}


\section{Introduction}

AI has become so adept at understanding complex requirements and generating functional code that software engineers (SWEs), both students and professionals, are adopting it for programming assignments and production systems. Computing educators and workforce developers are discussing whether future SWEs will still need to (1) write code themselves, (2) be able to write code themselves, and (3) be able to understand and fix code themselves. It is the core premise of this paper that SWEs will still need to be able to understand and fix code themselves, especially their own code, such as students meeting learning requirements or demonstrating intellectual ownership, professionals contributing to critical software systems where AI cannot be responsible for their failures, and novice workers forming skills \cite{anthropic} and joining organizations' talent pipelines \cite{russinovich}.

While it could have been assumed that SWEs understand their own code and that the quality of their artifacts is a proxy to their level of understanding, those assumptions have become challenged as SWEs can now prompt a multitude of coding agents with their own tasks, generate code that passes tests, and submit it as their own work without reviewing the code or even the tasks themselves. Educators are having to reconsider grading strategies \cite{macneil2024}, when student-generated artifacts are inseparable from AI-generated artifacts. Proctored exams may suffice for some courses, but they are often impractical, and standardized exams cannot evaluate understandings of personal designs and implementations. Code reviews with managers can assure code quality, but they are resource-intensive for identifying potential gaps in understanding. Both educators and employers require a better method for evaluating SWEs' understanding of their own code.

We have designed and developed a method, called Personalized Probing Puzzles ($p^3$), for generating personalized assessments from personal artifacts, inspired by parsons puzzles which are code reconstruction tasks common in introductory programming courses. But unlike traditional parsons puzzles, $p^3$ is generated from personal artifacts, such as the source code submitted by students for their own programming assignments. One practical advantage of $p^3$ is that it can use existing test infrastructures despite the assessments being personalized.

We deployed $p^3$ in Carnegie Mellon University (CMU)'s cloud computing course, where we observed some students failing the puzzle despite it being generated from their own code, and some students solving the puzzle but reporting a lack of self-confidence. Future studies are required to correlate puzzle results with code understanding. The main aim behind $p^3$ is to investigate if it can help locating potential gaps in the SWEs' understanding of their code and their ability to fix it, so that they may recognize their training needs and calibrate their self-efficacy, instructors may use the tool to personalize training, and managers may use it to expedite code review.
\section{Background}

By generating personalized assessments from personal artifacts, $p^3$ addresses the problem of evaluating an individual in the context of their own artifacts. It differs from evaluating their artifacts as a proxy to evaluating the individual, which is undertaken by automated grading tools \cite{messer24} which cannot reliably discern between human- and AI-generated artifacts. It also differs from evaluating the individual without their artifacts, which measures their understanding of programming concepts and programming in general, but not their understanding of the program itself \cite{Heinonen23}. Such in-context understanding is essential for both academic integrity and responsible AI use in production, where critical systems must be serviced by individuals with not only subject matter and AI expertise, but also semantic understanding of the systems, including the systems' underlying assumptions.

Although the problem of needing to evaluate SWEs' understanding of their own code is exacerbated only recently, techniques from prior work can be applied such as "bebugging" \cite{bebugging}, which injects faults into programs in order to evaluate programmers' ability to find those faults. Recent work also investigated techniques for introducing realistic bugs \cite{patra21}. Hypothesizing that some students in computing education are using AI to the extent where they cannot reconstruct or recognize their own code, let alone detect bugs in it, we prototyped $p^3$ to test that hypothesis, with parsons puzzles as the assessment technique.

Parsons puzzles are a type of coding exercise that have been used
primarily in introductory programming courses \cite{syslit}. Solving a parsons
puzzle usually entails re-ordering (and sometimes re-indenting)
shuffled code blocks (i.e., segments of a program), in order to
construct the specified working program rather
than implementing it from scratch. A practical advantage of parsons puzzles is that they are much faster to solve than programming from scratch \cite{ericson17}, not only because the SWE does not have to type the code out, but also because realistic code often utilizes libraries and frameworks for which the SWE cannot be reasonably expected to memorize all syntax.

We can evaluate SWEs' understanding of their artifacts using other assessment techniques, including other parsons puzzles variations (Adaptive~\cite{adaptive},
Faded~\cite{faded}, and Two-Dimensional~\cite{2d}) and fault injection as discussed earlier, but they all rely on us having located segments of the artifacts against which to evaluate the SWEs' understanding, as we cannot expect them to reconstruct entire artifacts. The results presented in this study are based on puzzles generated from artifacts where we know the exact file paths and lines of code where understanding is important, but in order for $p^3$ to be viable for evaluating SWEs' understanding of open-ended designs and implementations, we need an algorithm for picking snippets of code (from their artifacts) that demonstrates specific competencies or functionalities. We will report on that algorithm in our future work.
\section{Design and Implementation}

\begin{figure}
\startmyfigspace
    \centering
    \includegraphics[width=1\linewidth]{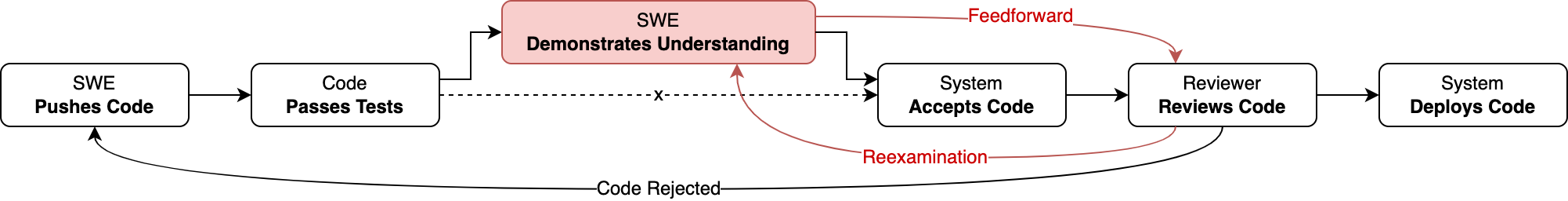}
\startmycapspace
    \caption{\sl SWE demonstrates understanding before system accepts their code.}
    \label{fig:workflow}
\myfigspaceend
\end{figure}

We designed $p^3$ to fit in a software development life cycle (Fig~\ref{fig:workflow}). After an SWE tests their code, they push it. Then, they demonstrate their understanding of the pushed code by solving personalized probing puzzles generated from that code before the system accepts it for review. Once the system accepts the code, the SWE can move on to their next task while the reviewer reviews it asynchronously; here the reviewer can be an instructor or a manager. $p^3$'s evaluation of the SWE's understanding (based on puzzle results) is forwarded to that reviewer, both for them to evaluate the SWE and for them to strategize their own review of the code. The reviewer may also want to re-evaluate the SWE by prompting them to demonstrate their understanding again with new puzzles.

\begin{figure}[t]
\startmyfigspace
    \centering
    \includegraphics[width=1\linewidth]{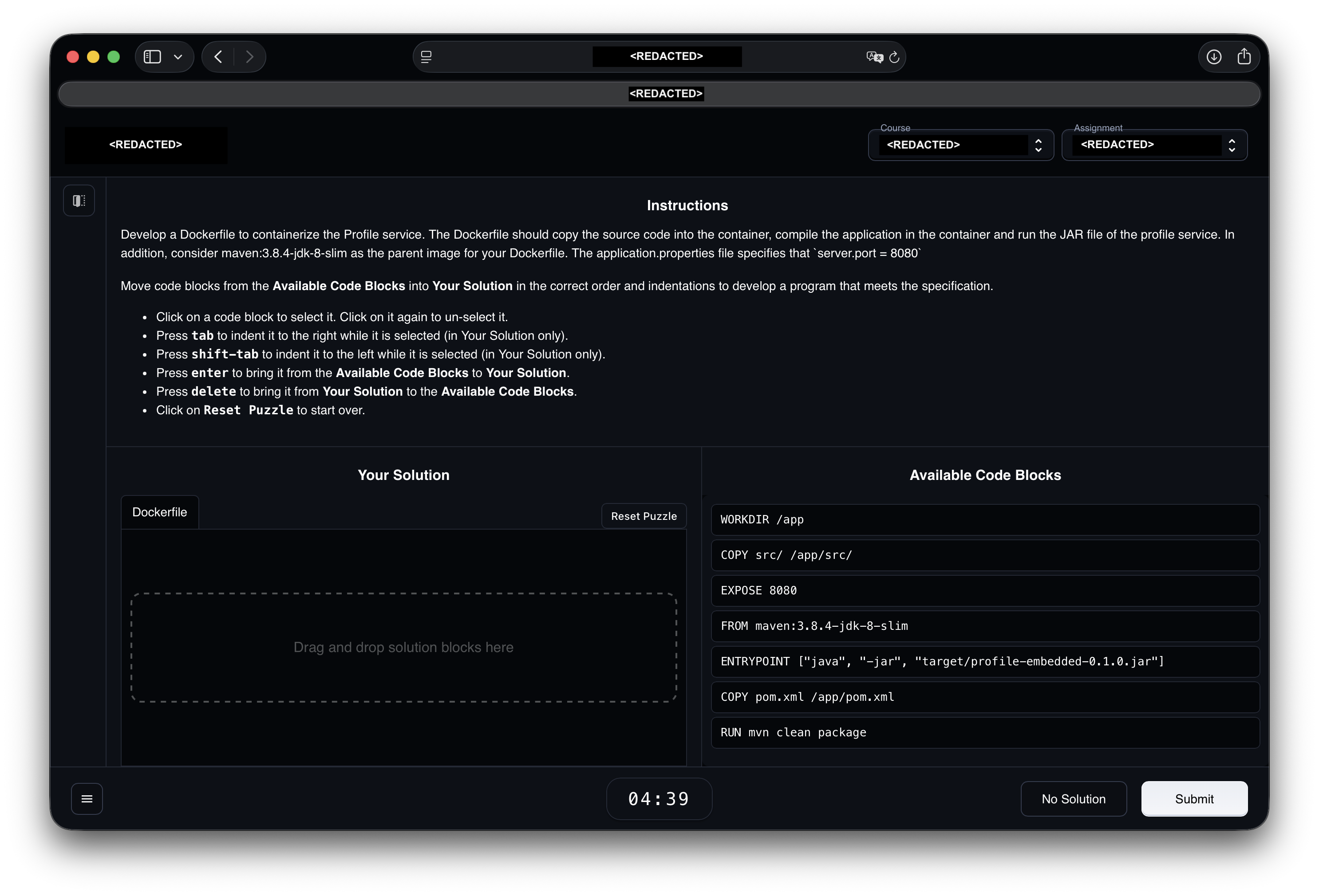}
\startmycapspace
    \caption{\sl The user interface of $p^3$, with a parsons puzzle of a Dockerfile}
    \label{fig:puzzle}
\myfigspaceend
\end{figure}

The current interface of $p^3$ is a single-page web application (Fig~\ref{fig:puzzle}). Users follow \textbf{Instructions} (center-top) to move \textbf{Available Code Blocks} (bottom-right) into their \textbf{Solution} (bottom-left), by dragging-and-dropping individual code blocks. There is a \textbf{Reset Puzzle} button to start over, a \textbf{Submit} button to submit their solution, a \textbf{No Solution} button for users who think their puzzle cannot be solved with the Available Code Blocks, and a \textbf{Timer} at the bottom. When the time is up, the puzzle is automatically submitted. Not all code blocks need to be used in the solution; some may be distractors \cite{adaptive}. Not all puzzles are solvable with the available code blocks; some code blocks may have been removed.


Behind the scenes, $p^3$ is also equipped with a \textbf{Snippet Picker} for picking and validating snippets from the source code (which may include multiple modified files and too many lines of code to fit in a single puzzle); a \textbf{Puzzle Maker} that compares puzzle-making strategies based on their efficacy in assessing SWE understanding of the same snippet; a \textbf{Puzzle Assigner} that assigns puzzles to SWEs, which can be fully automated based on their association with experimental groups; a  \textbf{Puzzle Logger} that tracks an SWE's every drag, drop, click, and keystroke within $p^3$ so reviewers can analyze the SWE's puzzle attempt; and a \textbf{Browser Logger} that tracks an SWE's every \texttt{window\_blur}, \texttt{window\_focus}, \texttt{tab\_hidden}, \texttt{tab\_visible}, \texttt{window\_resize} events as defined by Page Visibility APIs so reviewers can validate the integrity of the SWE's attempt.





\section{Pilot Study and Results}

\begin{listing}
\centering
\begin{Verbatim}[fontsize=\small]
FROM maven:3.8.4-jdk-8-slim
WORKDIR /app
COPY pom.xml /app/pom.xml
COPY src/ /app/src/
RUN mvn clean package
EXPOSE 8080
ENTRYPOINT ["java", "-jar", "target/profile-embedded-0.1.0.jar"]
\end{Verbatim}
\caption{\sl The reference solution Dockerfile has 7 lines of code}
\label{lst:dockerfile}
\end{listing}

$p^3$ was piloted in CMU's cloud computing course in Spring 2026 where students were presented a puzzle based on a programming task they had just completed. For that task, students write a Dockerfile to containerize a Java application. The task is one of seven that form a larger, two-week project on containerization. The task is simple, and a correct Dockerfile can be 7 lines long (Listing \ref{lst:dockerfile}). Students are expected to work independently on the task without peer help or AI assistance. They can check their solution against the auto-graders any number of times, and receive a score and feedback immediately based on the state of the Java application deployed from their Docker image. The puzzle itself, where students reconstruct the Dockerfile, must be completed in under 5 minutes, but students could start the puzzle anytime the day after the project was due. The puzzle was graded but the grade was released a week later. 159 students attempted the puzzle, 111 gave research consent; we report on the latter.

All 111 students received a full task score from the auto-graders. Group-Own (58/111) saw a puzzle of their own Dockerfile, Group-Ref (53/111) saw a puzzle of the reference Dockerfile which could be identical to their solution but often differs. All 111 students completed the post-puzzle survey.
\begin{itemize}
    \item 77.6\% of Group-Own and 66.0\% of Group-Ref solved their puzzle. 43.1\% of Group-Own and 35.8\% of Group-Ref reported high self-confidence and low effort in solving the puzzle in the post-puzzle survey. The numbers suggest about a third of students may not understand their own code, and more than half lack confidence in understanding their own code, which is striking.
    \item A handful of students who failed the puzzle admitted they did not write their own Dockerfile. A handful incorrectly suggested that their puzzle was impossible with the available code blocks, despite their puzzle being made of their own complete submissions. A handful reflected that attempting the puzzle prompted them to review their own code afterwards.
\end{itemize}

Although $p^3$ generates much more data than presented in this analysis, the study shows that $p^3$ can help locate potential gaps in SWEs' understanding of their own artifacts by generating personalized assessments from those artifacts. The assessments can be generated automatically, administered asynchronously, and finished in minutes. $p^3$ is designed to support additional assessment techniques other than parsons puzzles, and to gather additional behavioral metrics for interpreting puzzle results. Since the pilot study, $p^3$ has been deployed in 3 other courses at CMU (Python programming, compilers, and cloud native), and is scheduled to be deployed in a computer systems course, as we assess its efficacy across programming languages and domain expertise.
\section{Limitations and Future Work}

An assumption made in this study is that an SWE's ability to reconstruct or fix their program (as measured by the puzzle results) is good proxy to their understanding of that program. We plan to further validate that assumption in a future study, where that ability is measured by not only accuracy but also effort, self-efficacy, duration, and time passed since artifact creation, etc. A follow-up study is underway where code reviews are conducted where teaching assistants evaluate students' understanding of their own artifacts, although studies show inter-rater inconsistencies are common.

Another assumption made in this study is that no students failed the puzzle due to confusion with the user interface or user experience (UI/UX). A playground puzzle has since been developed so that future experiments start with SWEs familiarized with the UI/UX before they are evaluated using $p^3$.

\begin{credits}
\subsubsection{\ackname} 
This research is funded in part by the Carnegie Mellon-Accenture Center of Excellence in AI-Enabled Workforce Training (ACE-AI).

\end{credits}
%
%
%
\footnotesize
\bibliographystyle{seth}
\bibliography{refs}
\end{document}